\documentclass{ws-procs975x65}

\begin{document}

\title{CONSISTENT GRAVITATIONAL ANOMALIES FOR CHIRAL SCALARS}

\author{PIETRO MENOTTI}

\address{Department of Physics, University of Pisa,\\
and INFN Sezione di Pisa\\
Pisa, 56127, Italy\\
$^*$E-mail: menotti@df.unipi.it}
\begin{abstract}
Starting from the Henneaux-Teitelboim action for a chiral scalar, which
generalizes to curved space the Floreanini-Jackiw action, we give two simple
derivations of the exact consistent gravitational anomaly. The first
derivation is through the Schwinger-DeWitt regularization. The second exploits
cohomological methods and uses the fact that in dimension two the
diffeomorphism transformations are described by a single ghost which allows to
climb the cohomological chain in a unique way.
\end{abstract}


\bodymatter

\section{Gravitational anomalies}\label{aba:sec1}

The simplest instance of pure gravitational anomaly is the one due to the
presence of a chiral fermion in two dimensions (Ref.~\citen{AGW}).
There are also gravitational anomalies produced by boson fields i.e. by the
self-dual and anti self-dual fields which are realized in the simplest instance
by the chiral scalars in dimension $2$.
The coupling of (anti) self-dual
tensors to gravity is given by the Henneaux-Teitelboim action (Ref.~\citen{HT})
which generalizes to curved backgrounds the Floreanini-Jackiw action 
(Ref.~\citen{FJ})
for chiral scalars in two dimensions. On curved background the
chirality condition becomes (Ref.~\citen{BvN})
$
E_+^{\ \mu}\partial_\mu \varphi = 0
$
where $E^\mu_+$ are the inverse zweibeins.
In the following the key role will be played by the adimensional function
\begin{equation}
K= \frac{E_+^{\ 1}}{E_+^{\ 0}}=\frac{N}{\sqrt{h}}-N^1 =
\frac{\sqrt{-g}-g_{01}}{g_{11}}.
\end{equation}
The action is provided by (Ref.~\citen{HT})
\begin{equation}\label{action}
S= -\frac{1}{2}\int{d^2x ~\partial_1 \varphi (\partial_0 \varphi +
K \partial_1 \varphi)}=\frac{1}{2}\int{d^2 x ~\varphi \partial_1(\partial_0+
K\partial_1)\varphi}.
\end{equation}
The lack of explicit invariance under diffeomorphisms is the origin of the
anomaly.
The variation of $S$ w.r.t $\varphi$ gives the equation of motion
\begin{equation}
\partial_1(\partial_0+ K \partial_1)\varphi=0
\end{equation}
and the action vanishes on the equation of motion.
Under an infinitesimal diffeomorphism $x^\mu\rightarrow x^\mu+\xi^\mu$ 
the transformation of $K$ and $\varphi$ are
$ 
\delta_\xi K = -\partial_0\Xi - \partial_1\Xi ~K + \Xi~\partial_1K
$
and $\delta_\xi\varphi=\Xi\partial_1\varphi$, with
$\Xi=\xi^1-K\xi^0$. Action (\ref{action}) is invariant (Ref.~\citen{BvN}) 
under such transformation.
Thus only one combination of $\xi^1,\xi^0$ enters the 
transformation (Ref.~\citen{becchi}).
\section{The exact consistent anomaly through Schwinger DeWitt} 
The generating functional is given by
\begin{equation}\label{Z}
 Z[K] =e^{iW[K]} =  \int{{\cal D}[\phi] \exp\left[i\frac{1}{2}\int{d^2x \phi
(\partial_1 \partial_0  + \partial_1 K \partial_1) \phi}\right]}
\equiv  \left(\det (-i H )\right)^{-\frac{1}{2}} .
\end{equation}
The anomaly is provided by the variation of (\ref{Z}) under an infinitesimal
diffeomorphisms. We have for the variation $i\delta_\xi W[K]$ the following
expression
\begin{equation}
\int {\cal D}[\phi] e^{\frac{i}{2}\int
d^2x\phi(\partial_1(K\partial_1+\partial_0)\phi} \int
d^2x\frac{i}{2}\phi\partial_1(\delta_\xi K \partial_1\phi)/Z[K]
=\frac{1}{2}\int d^2x ~\delta_\xi H ~G(x,x')|_{x'=x}
\end{equation}
with
$G(x,x')$ is the exact Green function in the external field $K$, which will be
regularized \`a la Schwinger-DeWitt
\begin{equation}
G(x,x',\varepsilon)=i\langle x|\int_\varepsilon^\infty
e^{iHt}dt|x'\rangle~~~~{\rm while}~~~~\delta_\xi H=\partial_1(\Xi H) -
H\Xi\partial_1. 
\end{equation}
Taking into account that $H$ is the Laplace-Beltrami operator in the metric 
$g_{11}=0,~g_{10}=g_{01}=2,~g_{00}=-4K$ we obtain (Ref.~\citen{GM}) using the 
Seeley-DeWitt technique 
\begin{equation}\label{anomaly}
\delta_\xi W = -\frac{1}{24\pi}\int d^2x~\Xi(x)~\partial_1^3K(x)=
\frac{1}{24\pi}\int d^2x ~K(x)~\partial_1^3\Xi(x)\equiv G^E[K,\Xi]
\end{equation}
which is the Einstein anomaly.

\section{\bf W-Z consistency condition and non triviality of 
the anomaly}
Introducing the anticommuting diffeomorphism ghosts $v^1, v^2$ 
the BRST variation of $K$ is
\begin{equation}\label{deltaK}
\delta K= -\partial_0 V-\partial_1 V K + V\partial_1 K
\end{equation}
where $V= v^1-Kv^0$,  it is possible to show (Ref.~\citen{GM}) that 
(\ref{anomaly}) satisfies the
Wess-Zumino consistency relation $\delta G^E[K,V]=0$ and that the anomaly is
not trivial. I.e. with $Q^1_2= V\partial_1^3Kdx^1\wedge dx^0$ and
$\delta Q^1_2 = -d Q^2_1$, $\delta Q^2_1 = -d Q^3_0 =
-\frac{1}{2}d (V\partial_1V\partial^2V)\equiv -d N^3_0$, 
we have $Q^3_0\neq \delta X^2_0$ for any $X^2_0$.

\section{\bf Cohomological derivation of the anomaly}\label{cohosection}
Very general cohomological treatments of anomalies for conformal invariant
theories have been given (Ref.~\citen{BDK,BTV}) using two ghosts. Here 
exploiting the
fact that in our 
case we can work with a single ghost (Ref.~\citen{becchi}) the cohomological 
procedure can be streamlined.
In the following discussion we shall denote by $S^n(m)$ the space of terms
containing $n$ ghosts and $m$ derivatives, e.g. $V\partial_1^2V \partial_0 K
f(K)\in S^2(3)$.
The uniqueness of the last term $Q^3_0$ in the cohomological chain is 
equivalent to proving  that the sequence
\begin{equation}\label{S00sequence}
S^0(0) \stackrel{\delta}{\rightarrow} S^1(1)
\stackrel{\delta}{\rightarrow} S^2(2) \stackrel{\delta}{\rightarrow}
S^3(3) \stackrel{\delta}{\rightarrow} S^4(4)
\stackrel{\delta}{\rightarrow} 0
\end{equation}
differs from an exact sequence only in the penultimate junction due to the
presence of the non trivial term $N^3_0\equiv {\rm const}~ 
V\partial_1V\partial^2_1V$.
A great simplification (Ref.~\citen{GM}) is obtained by performing a change of 
basis
replacing the basis element $\partial_0V$ by $W\equiv \delta K\equiv
 -\partial_0 V-\partial_1 V K + V\partial_1 K$, which is
equivalent to it due to the relation (\ref{deltaK}). Then the algebra we
need to  use is simply $\delta V=V\partial_1V;~\delta K=W;~\delta W=0$.
The sequence (\ref{S00sequence}) is shown in Fig.1 where the numbers
on the vertical bars denote the dimension of the space.
The next step is to show the uniqueness of the term $Q^2_1$ 
up to trivial additions.
To this end we have to prove that the kernel of $\delta$ from
$S^2(3)$ into $S^3(4)$ is zero, modulo the trivial terms
$\delta S^1(2)$.
To this end we show (Ref.~\citen{GM}) that the sequence
\begin{equation}\label{S01sequence}
0 \stackrel{\delta}{\rightarrow} S^0(1) \stackrel{\delta}{\rightarrow}
S^1(2) \stackrel{\delta}{\rightarrow} S^2(3)
\stackrel{\delta}{\rightarrow} S^3(4)
\end{equation}
is exact.

We are left now with climbing the last step of the cohomology chain i.e. 
we have to prove the uniqueness, up to trivial terms, of the solution of
\begin{equation}
\delta Q^1_2= -d Q^2_1
\end{equation}
of which we know already a solution i.e. $Q^1_2 = {\rm const}~
V\partial_1^3 K dx^1\wedge dx^0$. 
It corresponds to proving the exactness of
sequence
\begin{equation}\label{S02sequence}
0 \stackrel{\delta}{\rightarrow} S^0(2) \stackrel{\delta}{\rightarrow}
S^1(3) \stackrel{\delta}{\rightarrow} S^2(4).
\end{equation}
which can be easily performed using the algebra and the change of basis 
described after eq.(\ref{S00sequence}). 
\begin{figure}
\psfig{file=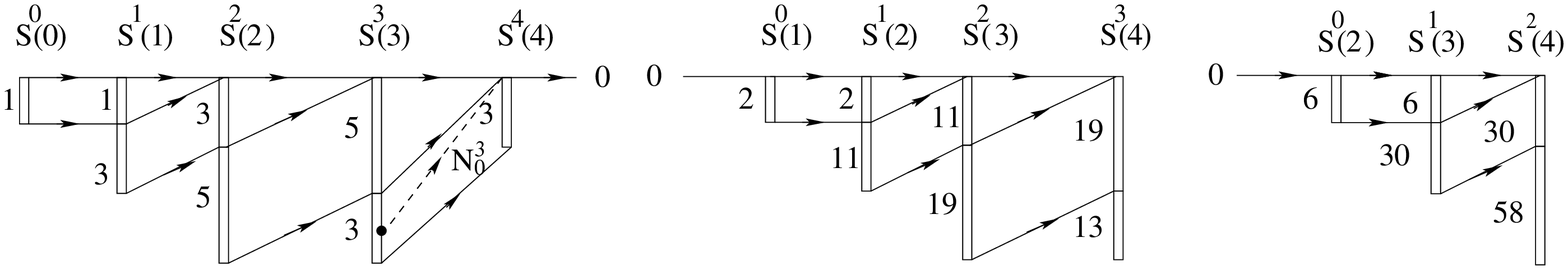,width=5in}
\caption{The three cohomological sequences}
\label{aba:fig1}
\end{figure}

\end{document}